\documentclass[galaxies,review,accept,moreauthors,pdftex]{mdpi} 

\firstpage{1} 
\pubvolume{xx}
\issuenum{1}
\articlenumber{5}
\pubyear{2019}
\copyrightyear{2019}
\history{Received: date; Accepted: date; Published: date}

\continuouspages{yes}

\newcommand{\Lar}{r_{\rm L}}
\newcommand{\lb}{l_b} 

\renewcommand{\vec}[1]{\mathbf{#1}}	
\newcommand{\dd}{\mathrm{d}}        

\newcommand\oderiv[2]{\displaystyle\frac{\dd #1}{\dd #2} }

\newcommand{\cm}{\,{\rm cm}}    
\newcommand{\pc}{\,{\rm pc}}     
\newcommand{\kpc}{\,{\rm kpc}}  

\newcommand{\GHz}{\,{\rm GHz}}  
\newcommand{\MHz}{\, {\rm MHz}} 
\newcommand{\s}{\,{\rm s}}      
\newcommand{\yr}{\,{\rm yr}}    
\newcommand{\Myr}{\,{\rm Myr}} 

\newcommand{\mkG}{\,\mu{\rm G}} 

\newcommand{\dyn}{\,{\rm dyn}}  
\newcommand{\erg}{\,{\rm erg}}  
\newcommand{\eV}{\,{\rm eV}}  
\newcommand{\GeV}{\,{\rm GeV}}  

\newcommand{\brms}{\,b_{\rm rms}}

\newcommand{\ncr}{n_{\rm cr}}

\newcommand{\kf}{k_\mathrm{F}}

\newcommand{\HI}{\mathrm{H\,\scriptstyle I}}

\newcommand{\uvec}[1]{\boldsymbol{\hat{\textbf{#1}}}}
\newcommand\Eq[1]{Eq.~\ref{#1}}
\newcommand\Fig[1]{Fig.~\ref{#1}}
\newcommand\Sec[1]{Section~\ref{#1}}
\Title{Revisiting the Equipartition Assumption in Star-forming Galaxies}

\Author{Amit Seta $^{1,2}$ and Rainer Beck $^{3}$}

\AuthorNames{Amit Seta and Rainer Beck}

\address{%
$^{1}$ \quad School of Mathematics, Statistics and Physics, Newcastle University, Newcastle Upon Tyne, NE1 7RU, UK; \\
$^{2}$ \quad Research School of Astronomy and Astrophysics, Australian National University, Canberra, ACT, Australia; amit.seta@anu.edu.au \\
$^{3}$ \quad Max-Planck-Institut fu\"{r} Radioastronomie, Auf dem H\"{u}gel 69, 53121 Bonn, Germany; rbeck@mpifr-bonn.mpg.de}

\corres{Correspondence: amit.seta@anu.edu.au} 

\abstract{Energy equipartition between cosmic rays and magnetic fields is often assumed to infer magnetic field properties from the synchrotron observations of star-forming galaxies. However, there is no
compelling physical reason to expect the same. We aim to explore the validity of the energy equipartition assumption. After describing popular arguments in favour of the assumption,
we first discuss observational results which support it at large scales and how certain observations show significant deviations from equipartition 
at scales smaller than $\approx 1 \, {\rm kpc}$, probably related to the propagation length of the cosmic rays. Then we test the energy equipartition
assumption using test-particle and MHD simulations. From the results of the simulations, we find that the energy equipartition assumption is not valid at scales smaller than the driving scale of the
ISM turbulence ($\approx 100 \, {\rm pc}$ in spiral galaxies), which can be regarded as the lower limit for the scale beyond which equipartition is valid. 
We suggest that one must be aware of the dynamical scales in the system before
assuming energy equipartition to extract magnetic field information from synchrotron observations. 
Finally, we present ideas for future observations and simulations to investigate in more detail under which conditions the equipartition assumption is valid or not.}


\begin{document}

\section{Introduction}
\label{sec:intro}

Magnetic fields are an important component of the interstellar medium (ISM) of star-forming galaxies. They provide additional pressure support to the gas against the gravitational field \citep{BC90}, 
control the propagation of cosmic rays \citep{KP1969,Wentzel1974,Zweibel13,SSSBW17}, affect the star formation rate \citep{BBT15,KF2019}, suppress galactic outflows \citep{Bendre15,Shukurov18}, and 
also affect the properties of the ISM phases \citep{Shukurov18}. Their role in the formation and evolution of galaxies is still not known. There are a number of observational tracers of galactic magnetic fields \citep{Klein2015}: 
synchrotron emission, Faraday rotation, optical polarization, dust polarization, and the  Zeeman effect. The existence of galactic magnetic fields is usually explained by the turbulent dynamo theory \citep{Beck1996,BS2005}, 
a process by which the kinetic energy of the turbulent ISM is converted into the magnetic field energy. 
The turbulence in spiral galaxies is driven by supernova explosions, which stirs the ISM at length scales approximately within the range
$50 \text{--} 100 \pc$ \footnote{Even smaller values of the driving scale of turbulence in spiral galaxies, $1 \text{--} 20 \pc$, are also reported \citep{MS96,Haverkorn08,Iacobelli2013}. 
The driving scale of turbulence can also be greater than $100 \pc$ when energy injection by superbubbles is considered (Luke~Chamandy, priv. comm.).} 
\citep{OS93,Gaenslar2005,Fletcher2011,Houde2013}. This scale is referred to as the driving scale of the turbulence, $l_0$. Based on $l_0$, the galactic magnetic field can be 
divided into the small- and large-scale (or equivalently turbulent and regular) components. The correlation length of small-scale magnetic fields, referred to as $\lb$, is comparable to $l_0$ and that of the large-scale field
is much larger than $l_0$ (few $\rm kpc$ in spiral galaxies). The dynamo theory is also conventionally divided into two parts: the small-scale or fluctuation dynamo theory to explain the small-scale field
and the $\alpha-\Omega$ dynamo (or mean field dynamo) theory to describe the large-scale field. It is important to observationally study the properties of magnetic fields in galaxies to characterize their role in various galactic processes 
and to better understand the physics of the turbulent dynamo theory.

Synchrotron emission is one of the most powerful probes of magnetic fields in spiral galaxies. The intensity of total synchrotron emission $I_\mathrm{syn}$ is a measure of the number density of cosmic-ray electrons (CREs) in the relevant energy range and of the strength of the total (sum of both the large- and small- scale components) magnetic field component $B_\mathrm{tot,\perp}$ perpendicular to the line of sight (i.e. in the sky plane). 
The synchrotron intensity, $I_\mathrm{syn}$, is given by
\begin{equation} \label{syn}
I_\mathrm{syn} = C \int_L N_\mathrm{CRE} \, B_\mathrm{tot,\perp}^{\,\,\,(\epsilon+1)/2} \,\, \dd\ell\,,
\end{equation}
where $C$ is a constant, $N_\mathrm{CRE}$ is the number density of CREs,  $L$ is the total path length, and $\epsilon$ is the power-law index of CREs energy spectrum.
Radio continuum emission from star-forming galaxies is a mixture of thermal and non-thermal (synchrotron) components. The non-thermal fraction dominates at radio wavelengths of more than a few cm. 

To measure $B_\mathrm{tot,\perp}$ from $I_\mathrm{syn}$, one needs independent information about the density and spectrum of CREs, which is available only in the Milky Way and a few external galaxies (Section~\ref{sec:strengths}). For most external galaxies, the only presently available method is to assume equipartition between the energy densities of the total cosmic rays (dominated by protons) and the total magnetic field.
The magnetic field strength obtained by using the equipartition assumption is very similar
to that obtained from the minimum energy estimate (in this method the total energy of cosmic rays and magnetic fields is minimized to produce the observed synchrotron radiation) \citep{Longair1994,Klein2015}.
If energy equipartition between cosmic rays and magnetic fields is assumed, two further assumptions are required since most of the cosmic ray energy is due to protons
but the synchrotron emission is largely produced by electrons. Firstly, it is assumed that the cosmic ray protons  have a similar spatial distribution as cosmic ray electrons. 
Secondly, the ratio of the number of protons to the number of electrons is assumed to be a fixed constant,  $K$. Then the properties of magnetic field
can be estimated from the synchrotron observations. In this review, we focus on the energy equipartition assumption. 

We first present popular reasons to expect energy equipartition between cosmic rays and magnetic fields in Section~\ref{sec:why}. Then we describe the method to extract magnetic field strength from synchrotron observations using the energy equipartition assumption in Section~\ref{sec:equi}.  In Section ~\ref{sec:obs}, we summarize the pro and con arguments on the equipartition assumption from observations. Then we present new numerical simulations of the interaction between magnetic fields and cosmic rays in Section~\ref{sec:cr}. Finally, we conclude in Section~\ref{sec:con} and suggest some future directions of research.

\section{Cosmic rays and magnetic fields: why do we expect energy equipartition?}
\label{sec:why}

The following two arguments are used to justify energy equipartition between cosmic rays and magnetic fields. 
First, both the cosmic rays and magnetic fields have a common source of energy (supernova explosions) 
and thus, in an energy equilibrium state, they would equally share the total energy from the source. 
Cosmic rays are accelerated in supernova shocks \citep{BellI78, BellII78, BO78, Drury83}. Supernovae are also a major driver of the ISM turbulence, which 
amplifies galactic magnetic fields. Thus, cosmic ray and magnetic field energies are derived from a single source
and over large scales in length and time, both components would have roughly equal energy. This only applies to large scales
where a dynamic equilibrium between the two components can be considered. This is also probably true only for systems
where an equilibrium has time to be established, i.e., galaxies at the present epoch. It is unlikely that the equipartition assumption 
holds for violently active systems such as young galaxies and starburst galaxies.
However, the equipartition assumption
is widely used to obtain magnetic field strengths from synchrotron intensity, independent of the spatial resolution of
observations and the system under consideration. An another version of this argument 
implies local pressure equality between cosmic rays and magnetic fields.
The second argument for equipartition is 
that the cosmic rays are confined by magnetic fields and thus a correlation between them is expected 
\citep{B56,Stepanov2009,Stepanov2014}. Both these arguments may sound convincing but are not completely compelling. 
There are a few observational signatures in support of the energy equipartition at larger scales ($\ge \kpc$) in the ISM of star-forming galaxies (discussed in \Sec{sec:obs}),
but the detailed physics of each argument is yet to be explored. Thus, it is important to revisit the energy equipartition assumption and test its validity.

\section{The equipartition method}
\label{sec:equi}

\subsection{Basic assumptions}

\noindent The cosmic ray number density $N(E)$, where $E$ is the particle energy, has a power-law energy spectrum 
\begin{equation}
N(E) \propto E^{-\epsilon} \,.
\label{eq:spectrum}
\end{equation}
The radio synchrotron intensity also follows a power law, $I_\mathrm{syn} \propto \nu^{-\alpha}$, where $\alpha=(\epsilon-1)/2$ is the spectral index.

The total cosmic-ray energy density is determined by integrating over their power-law energy spectrum.
For any mechanism of cosmic-ray acceleration that generates a power law in particle momentum, like diffusive shock acceleration (e.g. \cite{Bell1978a}), the transformation from the momentum spectrum to the energy spectrum causes a break in the energy spectrum at the rest mass energy, the transition from non-relativistic to relativistic energies. The spectrum at non-relativistic energies is flatter, with a spectral index of $-(\epsilon+1)/2$ \cite{Bell1978b}. The highest contribution to the total energy comes from particles just beyond the rest mass energy, while the contribution of non-relativistic particles is small, and the low-energy integration limit hardly affects the total energy density.

In the interstellar medium (ISM), most cosmic-ray particles are protons and electrons. Their different rest masses lead to different energy spectra. Beyond the proton rest mass of 938\,MeV, the energy spectra have the same slope $\epsilon$, but the number densities per energy interval are offset by the factor 
$K = (m_\mathrm{p}/m_\mathrm{e})^{(\epsilon - 1)/2}$ \cite{Bell1978b,Pohl1993}.
\footnote{Arbutina et al. \cite{Arbutina12} argue that $K$ is smaller if the injection energy is comparable to or larger than the electron's rest mass energy.}
For strong shocks and an adiabatic index of 5/3, the energy spectrum is predicted to have an initial (injection) spectral index of $\epsilon=2.0-2.4$ (Fig.~1 in \cite{Caprioli11}), which is consistent with the $\gamma$-ray emission observed from young supernova remnants (Fig.~3 in \cite{Caprioli11}). The typical injection spectral index $\epsilon\simeq2.2$, such as in the remnant of Tycho's supernova \cite{Morlino12}, yields $K\simeq 100$. Such a value is indeed derived at GeV energies from radio and $\gamma$-ray observations in the local Milky Way (compare Figs.~2 and 3 in \cite{Strong2004}) and also from direct CR measurements \cite{Picozza13}.

Making use of these properties, Beck \& Krause \cite{Beck2005} proposed a revised method to calculate the total magnetic field strength from the total synchrotron intensity $I_\mathrm{syn}$ that scales with the total magnetic field strength $B_\mathrm{tot,\perp}$ to the power $(3+\alpha)$, so that
$B_\mathrm{tot,\perp}$ scales as:
\begin{equation}
B_\mathrm{tot,\perp} \propto (I_\mathrm{syn} \, (K + 1) \, / \, L)^{\, 1/(3 + \alpha)} \, ,
\label{eq:equi}
\end{equation}
where $\alpha$ is the synchrotron spectral index and $L$ is the effective pathlength through the source. $K$ is the ratio of number densities of CR protons and electrons in the relevant energy range of a few GeV. 
$K\simeq100$ is a reasonable assumption for diffusive shock acceleration of cosmic rays in galaxy disks (see above). This method is widely used to estimate field strengths in galaxies. Two refinements, i.e. replacing the sharp spectral break assumed in \cite{Beck2005} by a smooth transition and taking different ion species into account  \cite{Arbutina12}, do not significantly modify the results.

Note that the ``classical'' equipartition estimates often presented in textbooks (e.g. \cite{Klein2015,Longair1994}) and in earlier papers (e.g. \cite{Fitt1993}) were based on an integration of the synchrotron spectrum between fixed radio frequencies, which introduces an implicit dependence on the total magnetic field and leads to a fixed exponent of 2/7 instead of $1/(3+\alpha)$ in Eq.~(\ref{eq:equi}). Still, the difference between the classical and the revised estimates is small for typical CRE spectra and weak magnetic fields (see Fig.~1 in \cite{Beck2005}).

\subsection{Restrictions}
\label{sec:res}

In several cases, the assumptions of the method \cite{Beck2005} are not fulfilled, so that its application may lead to incorrect results:

(1) The injection spectral index of cosmic rays could be smaller than $\epsilon=2.0$ if the adiabatic index is smaller than 5/3 due to the pressure of the relativistic particles. For such spectra, the integration over the energy spectrum has to be restricted to a limited energy interval.

(2) CREs suffer from energy losses (synchrotron, inverse Compton, ionization and relativistic bremsstrahlung), especially in starburst regions or massive spiral arms where magnetic field strength, photon density, and gas density are high. As energy losses of aging CREs are much more severe than those of cosmic-ray protons, the ratio $K$ increases \cite{Pohl1993}. $K$ is also expected to increase with increasing distance from the injection sites of CREs, e.g. in the outer disks and halos of galaxies. Using the standard value $K=100$ instead of its real value $K'$ underestimates the total magnetic field by a factor of $(K'/K)^{1/(3 + \alpha)}$ in such regions \citep{Beck2005}.

(3) In dense gas, e.g. in starburst regions, secondary positrons and electrons may be responsible for most of the radio emission via pion decay. Notably, the ratio of protons to secondary electrons is also about 100 at typical radio wavelengths \citep{Lacki2013}.

(4) For an electron-positron plasma, e.g. in jets of radio galaxies, $K=0$ is valid. Here, method \cite{Beck2005} gives too small values because the low-energy integration limit (related to the proton's rest mass energy) is too high.\\

\noindent Further more fundamental restrictions are:

(5) Due to the highly non-linear dependence of $I_\mathrm{syn}$ on $B_\mathrm{tot,\perp}$,
the average equipartition value $B_\mathrm{tot,\perp}$ derived from synchrotron intensity is biased towards high field strengths and is an overestimate if $B_\mathrm{tot}$ varies along the line of sight or across the telescope beam. For $\alpha=1$ and the equipartition case, the overestimation factor $g$ of the total field
(see Appendix A in \cite{Stepanov2014}) is
\begin{equation}
g = (<B_\mathrm{tot,\perp}^{\,\,4}>)^{1/4} \, / <B_\mathrm{tot,\perp}> \,\, =
(1 + (8/3) Q^2 + (8/9) Q^4)^{1/4} \, ,
\label{over2}
\end{equation}
where $< \,\,\, >$ indicates averaging along the line of sight and across the beam.\\
$Q=(<\delta B_\mathrm{tot,\perp}^2>)^{1/2}\, / <B_\mathrm{tot,\perp}>$ is the amplitude of the field fluctuations relative to the mean field. For strong fluctuations, which implies $Q\approx1$, 
the calculation gives an overestimate of the field strength by a factor of $\approx1.5$.

(6) Young supernova remnants, the sources of cosmic ray particles, are rare in galaxies, so that the cosmic-ray energy density varies with time and fluctuates in space. Reaching the balance of energy equipartition needs time for the cosmic rays to diffuse and smooth out the fluctuations. Hence, equipartition cannot be expected to be valid at small scales (see Section~\ref{sec:dev}). The relation between cosmic rays and magnetic fields at smaller
scales as predicted from numerical simulations is discussed in detail in Section~\ref{sec:cr}.

\section{Inferences from observations}
\label{sec:obs}

\subsection{Definition of magnetic field components}

The total magnetic field is separated into a {\em regular (large-scale)}\ and a {\em turbulent (small-scale)}\ component.
Total synchrotron emission traces the total magnetic field perpendicular to the line of sight, while polarized synchrotron emission traces the {\em ordered}\ field perpendicular to the line of sight at the scale of the telescope beam. Anisotropic turbulent, anisotropic tangled, and regular field components all contribute to the ordered field observed in polarization.
The polarization angle (corrected for Faraday rotation) shows the field {\em orientation}, which is ambiguous by $180^\circ$, and hence is not sensitive to field reversals. Faraday rotation (and the longitudinal Zeeman effect) is sensitive to the {\em direction}\ of the field along the line of sight and hence can unambiguously trace regular fields.

\subsection{Equipartition estimates in galaxies}
\label{sec:equigal}

The average equipartition strength of the total fields derived from the total synchrotron intensity (using the classical estimate, scaled to $K=100$) for a sample of 146 late-type spiral galaxies is $B_\mathrm{tot,\perp}=11\pm 4\,\mu$G \cite{Fitt1993}. The sample of 74 spiral galaxies gave a similar value of $B_\mathrm{tot}=9\pm 2\,\mu$G \cite{Niklas1995}.
The total equipartition field strength in the Milky Way is about $10\,\mu$G at 5\,kpc galactocentric radius with a decline to about $4\,\mu$G at 15\,kpc radius (Fig.~\ref{fig:MW}).

\begin{figure*}[t]
\vspace*{7mm}
\begin{center}
\includegraphics[width=10cm]{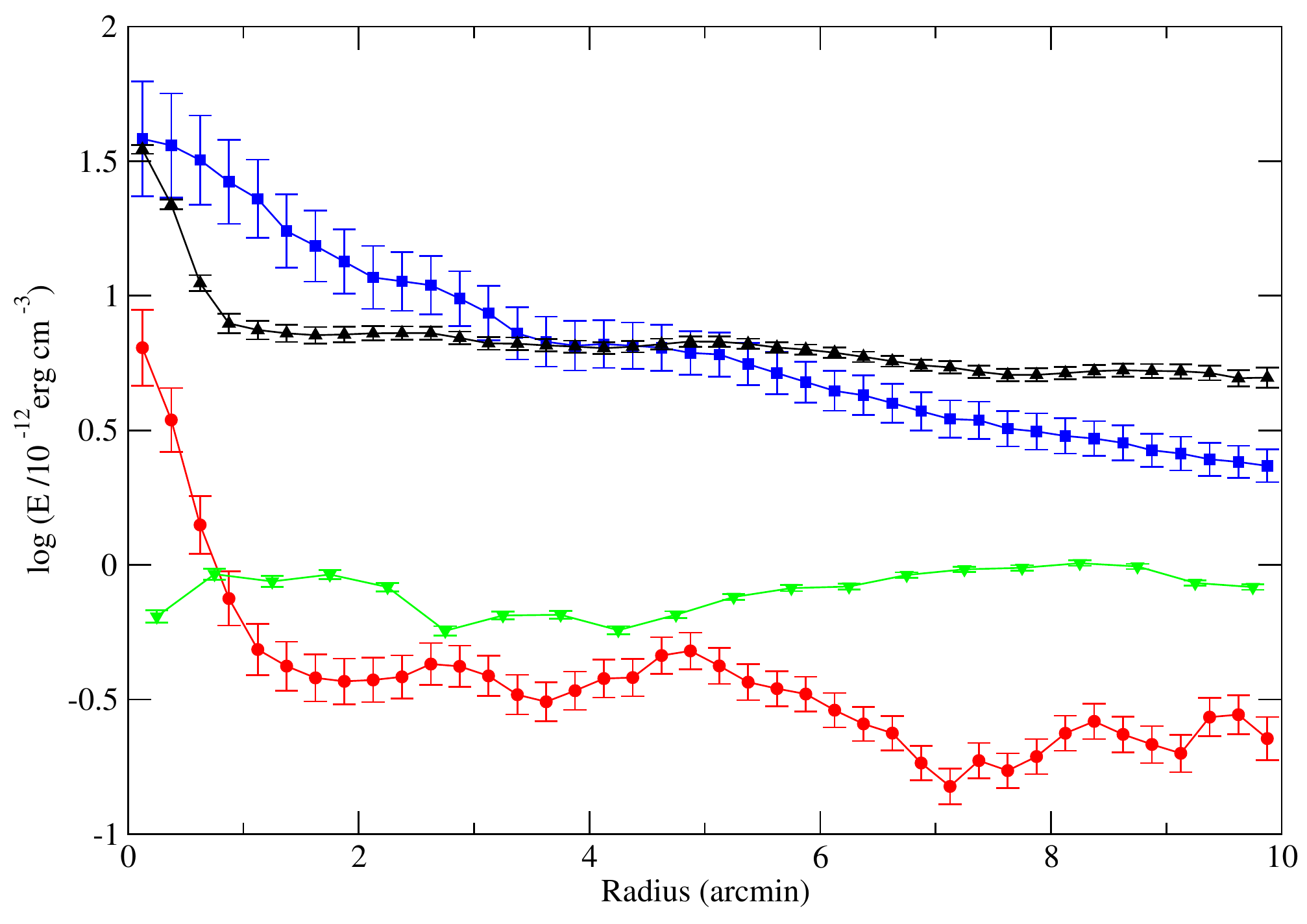}
\caption{Radial variation of the energy densities in IC\,342, determined from observations of synchrotron and thermal radio continuum and the CO and $\HI$ line emissions: magnetic energy density of the total field $B_\mathrm{tot}^2/8\pi$ (black triangles), identical to that of total cosmic rays; magnetic energy density of the ordered field $B_\mathrm{ord}^2/8\pi$ (green triangles); kinetic energy density of the turbulent neutral gas $E_\mathrm{turb}=0.5 \, \rho_n \, \mathrm{v_{turb}}^2$ (blue squares), assuming a turbulent velocity $\mathrm{v_{turb}}=10$~km/s, taken from the average velocity dispersion of the HI gas; and thermal energy density of the warm ionized gas $E_\mathrm{th}=1.5 \, n_\mathrm{e}\, k \, T_\mathrm{e}$ (red circles), where $n_\mathrm{e}$ is the average thermal electron density and $T_\mathrm{e}$ is the electron temperature ($T_\mathrm{e} \approx 10^4$~K). The error bars include only errors due to rms noise in the images from which the energy densities are derived. No systematic errors are included, e.g. those imposed by a radial variation of thermal gas temperature, filling factor or turbulent gas velocity, nor errors due to deviations from the equipartition assumption. At the distance of 3.5\,Mpc, 1\,arcmin corresponds to 1.0\,kpc (from \cite{Beck2015a}).}
\label{fig:ic342}
\end{center}
\end{figure*}

The equipartition estimates for several spiral galaxies \cite{Basu2013,Beck2015a} indicate that the magnetic energy density (dominated by the energy density of the small-scale field) is similar to the kinetic turbulent energy density (Fig.~\ref{fig:ic342}), as expected for the operation of a turbulent dynamo, while the thermal energy density is about one order of magnitude smaller (low-$\beta$ plasma).
A similar result was also obtained for the Milky Way in the solar neighborhood \cite{BC90}.

The magnetic energy density in IC\,342 seems to dominate at large radii (Fig.~\ref{fig:ic342}), which can be explained with non-linear dynamo models \citep{Chamandy2018} if the large-scale fields dominates out there, but this may also indicate that equipartition is no longer valid in the outer disk, e.g. due to fast cosmic-ray propagation.

\begin{figure*}[t]
\vspace*{7mm}
\begin{center}
\includegraphics[width=10cm]{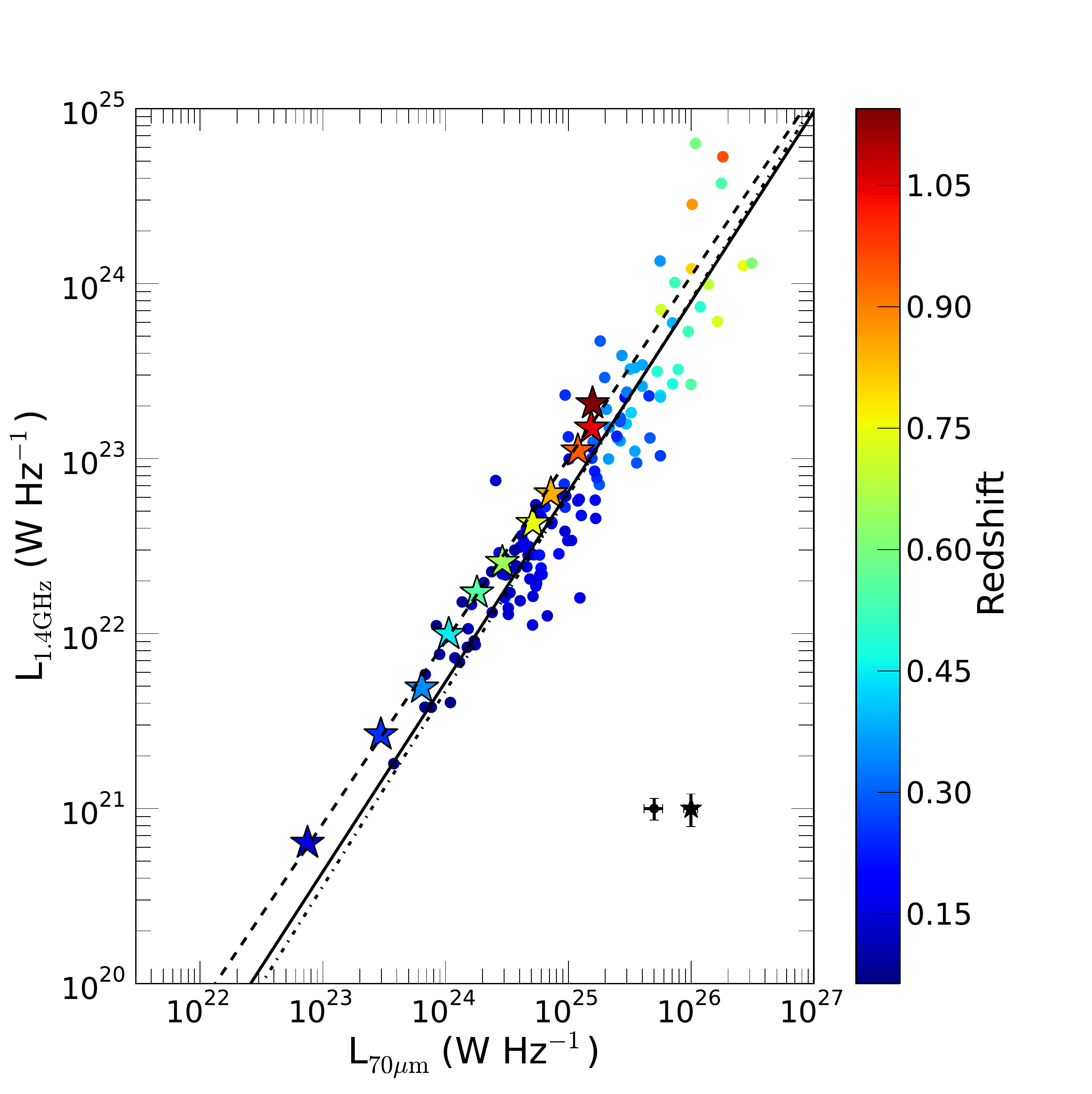}
\caption{Radio luminosity at 1.4\,GHz against monochromatic infrared luminosity at $70\,\mu$m at rest frames. Sources detected in the XMM-LSS field are shown as circles, stacked sources as stars. The symbols are color-coded based on their redshift. The solid line shows the fit to the entire data, the dashed and dashed–dotted lines are for the stacked and detected sources, respectively (from \cite{Basu2015}).}
\label{fig:RIC}
\end{center}
\end{figure*}

Another indication for the validity of the equipartition assumption at galaxy scale comes from the {\em global}\ correlation between the galaxy-integrated luminosity of the total radio continuum emission at frequencies of around 1\,GHz, which is mostly of synchrotron origin, and the infrared (IR) luminosity of star-forming galaxies. This is one of the tightest correlations known in astronomy. The correlation extends over five orders of magnitude \citep{Bell2003}, is slightly super-linear in log-log scale with an exponent
of $1.09\pm0.05$ \citep{Basu2015} (Fig.~\ref{fig:RIC}). The exponent of the non-thermal (synchrotron)--IR correlation must be steeper because the correlation between radio thermal and IR luminosities is linear \cite{Price1992} because UV photons ionize the gas and heat the warm dust. The exponent of the radio--IR correlation for thermal-corrected synchrotron luminosities is $1.33\pm0.10$ \cite{Price1992}. The equipartition model \cite{Niklas1997} relates total (mostly turbulent) magnetic fields, cosmic rays, gas density, and star formation, and is able to explain the super-linear synchrotron--IR correlation. \footnote{If the IR emission emerges mostly from the cool dust that is heated by the general interstellar radiation field, the exponent of the correlation can be smaller than 1 \cite{Hoernes98}.}

\subsection{How do magnetic field strengths derived from the equipartition assumption compare with those from other methods?}
\label{sec:strengths}

The Faraday rotation measure ($RM$) is a signature of the regular field $B_\mathrm{reg,\parallel}$ along the line of sight and can be used to constrain the total field strength in galaxies. For example, $|RM|\simeq$100\,rad/m$^2$ in the magnetic arms of the spiral galaxy NGC\,6946 yields $B_\mathrm{reg}\simeq8\,\mu$G for a typical electron density of $n_e\simeq0.03$\,cm$^{-3}$, a pathlength of 1\,kpc, and an inclination of the galaxy's disk of $30^\circ$ \cite{Beck2007}. A degree of field order of $B_\mathrm{reg}/B_\mathrm{tot}\simeq0.6$ follows from the observed degree of synchrotron polarization at high radio frequencies. As the contribution of anisotropic turbulent fields to the ordered field is probably small in the magnetic arms (located between the optical arms), $B_\mathrm{reg} \simeq B_\mathrm{ord}$. This gives $B_\mathrm{tot}\simeq13\,\mu$G, 
consistent with the equipartition estimate derived for the same regions.

Various measurements in the Milky Way confirm the equipartition estimate of the total field strength. From the dispersion of pulsar RMs, the Galactic magnetic field was found to have a significant turbulent component with a mean strength of about $6\,\mu$G \citep{Han2004}.
The strength of the local regular field is about $2\,\mu$G \citep{Han2006}. Zeeman splitting observations of low-density gas clouds (that have trapped the field from the diffuse ISM) yield field strengths of about $6\,\mu$G, corrected for effects of the line-of-sight \cite{Crutcher2010,Crutcher2012}.

The Voyager~1 spacecraft reached interstellar space in 2012 and since then measured a constant total field strength of $4.8\pm0.4\,\mu$G \citep{Burlaga2016}. The interstellar field is draped around the heliopause. A 3D magnetohydrodynamic (MHD) model of this interaction yielded a strength of the pristine local interstellar field of $2.9\pm0.1\,\mu$G \cite{Zirnstein2016}. This value is close to the strength of the local regular field.

Measurement of $B_\mathrm{tot}$ without the need of the equipartition assumption is possible if independent information about the cosmic-ray spectra is available. Cosmic rays can be measured directly in the solar neighborhood where equipartition was found to be valid and the magnetic field, cosmic ray, and kinetic energies are roughly equal \cite{BC90}. The density of CREs can be inferred from their X-ray emission by the inverse Compton (IC) effect, e.g. in galaxy clusters \cite{Govoni2004}. IC X-ray emission from star-forming galaxies has not yet been detected because other sources of X-ray emission are dominant. $\gamma$-rays generated by interactions of cosmic-ray protons with gas nuclei give information about the energy density of protons. Modeling the $\gamma$-ray and radio emissions allowed constraining the magnetic and cosmic-ray energy densities in the Milky Way and a few galaxies. Global equipartition is found to be valid in the Milky Way, M\,31, and in the LMC \cite{Yoast2016}.

\begin{figure*}[t]
\vspace*{7mm}
\begin{center}
\includegraphics[width=10cm]{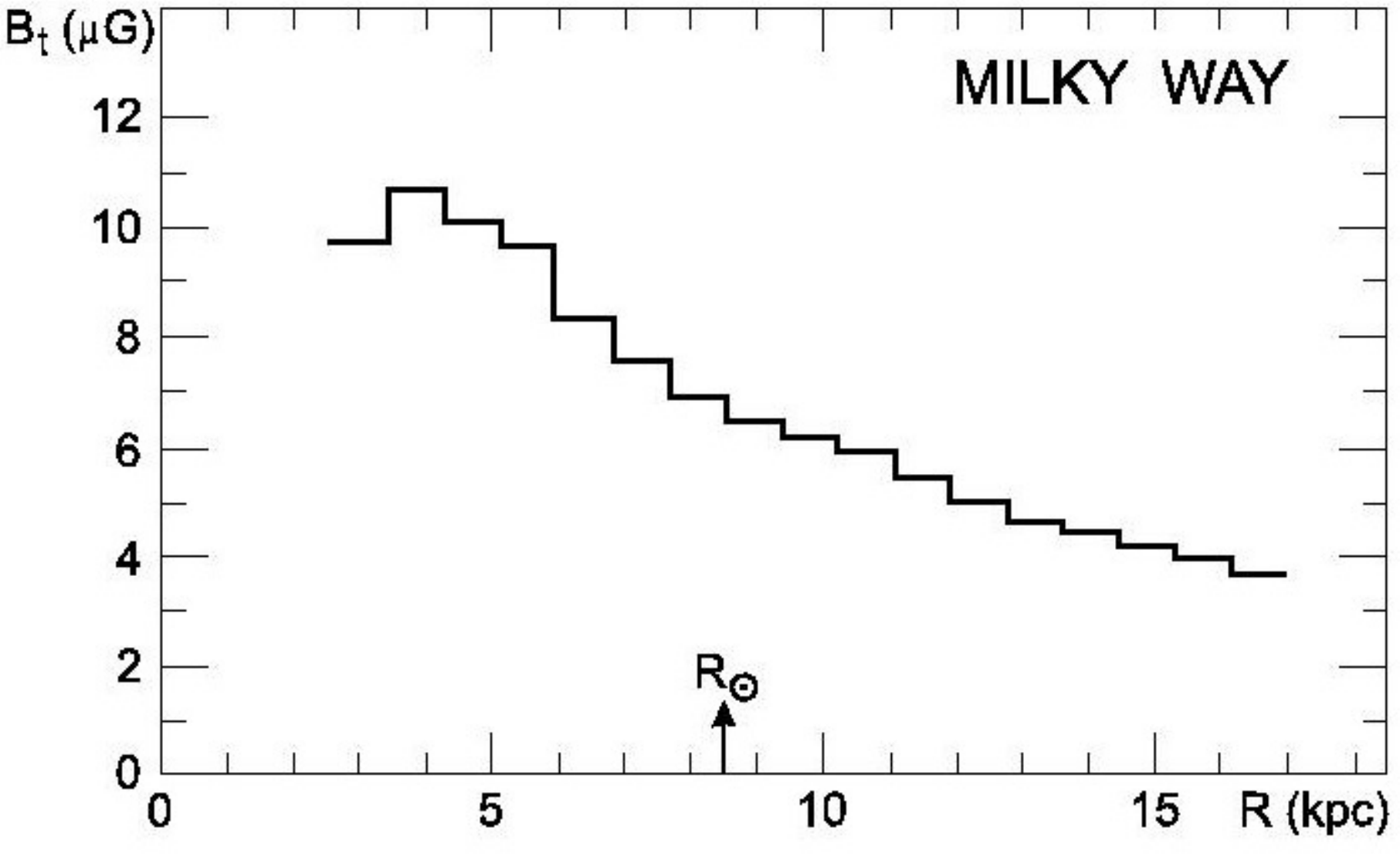}
\caption{Radial variation of the total magnetic field in the Milky Way, estimated with the equipartition assumption from the data of the all-sky radio survey at 408\,MHz \cite{Beuermann1985}. The proton/electron ratio was assumed to be $K=100$, the synchrotron spectral index $\alpha=1.1$ for 2.5\,kpc$<$R$<$ 6\,kpc and $\alpha=0.85$ for $R\ge6$\,kpc (from E.M.~Berkhuijsen, priv. comm.).}
\label{fig:MW}
\end{center}
\end{figure*}

From radio and $\gamma$-ray data in the Milky Way, a model of the radial variation of $B_\mathrm{tot}$ was constructed (Fig.~6 in \cite{Strong2000}) which agrees well with the equipartition estimates (Fig.~\ref{fig:MW}). More recent modeling yielded field strengths near the Sun of about 5\,$\mu$G of the isotropic turbulent field, about 2\,$\mu$G of the anisotropic turbulent field, and about 2\,$\mu$G of the regular field \cite{Orlando2013}, adding up (quadratically) to a total field strength of about 6\,$\mu$G.

\subsection{Observational indications for deviations from the equipartition assumption}
\label{sec:dev}

The total radio and IR intensities {\em within}\ galaxies are also highly correlated, but sub-linearly with exponents of $0.78\pm0.02$ for M\,31 and $0.46\pm0.02$ for M\,33, probably due to CRE propagation \cite{Berkhuijsen2013}. This indicates that equipartition between magnetic fields and cosmic rays is no longer valid at spatial scales below a few kpc. The exponent of the local correlation within 12 spiral galaxies varies between 0.3 and 0.8 \cite{Heesen2014}, probably due to differences in the mean propagation length (which is defined as the product of the lifetime within the galaxy and the mean propagation speed) 
of CREs due to diffusion or advection. For a constant density of CREs, the exponent of the local radio synchrotron--IR correlation is smaller by a factor of $(1+\alpha)/(3+\alpha)~\simeq0.4-0.5$ compared to that of the global correlation for which equipartition holds, consistent with the observations. As the lifetime of cosmic-ray protons is longer than that of electrons, their mean propagation length is also larger, which may increase the spatial scale beyond which equipartition is valid.

The correlation analysis of the fluctuations in total radio intensity observed in the Milky Way and the nearby galaxy M\,33 by \cite{Stepanov2014} gave further evidence that equipartition does not hold at scales smaller than about 1\,kpc. The equipartition field strength is probably underestimated in regions close to the CRE sources and overestimated in regions far away from the sources.

Modeling the radio and $\gamma$-ray data indicated that the global magnetic energy density is by a factor of several higher than the global cosmic-ray energy density in the central regions of the starburst galaxies M\,82 and NGC\,253, and even higher by several orders of magnitude in the ultraluminous infrared starburst galaxy Arp\,220 where a field strength on milligauss level was found \cite{Yoast2016}. It seems that the amount of deviation from equipartition increases with gas density, due to severe collisional energy losses of the cosmic-ray protons. Hence, the equipartition assumption heavily underestimates the magnetic field strength in dense starburst galaxies, which is of particular importance for the interpretation of galaxies in the early Universe.

The equipartition condition probably fails also in dwarf galaxies with a low star-formation rate (SFR) and hence a discontinuous supply of CRs \cite{Schleicher2016}. Below $SFR\simeq0.01$\, M$_\mathrm{o}$/yr, radio synchrotron emission does no longer trace the SFR and the equipartition field strength is no longer correlated with SFR \cite{Filho2019}.

\section{Testing of the equipartition assumption at small scales from direct numerical simulations}
\label{sec:cr}

Cosmic rays interact with magnetic fields in the ISM (further discussed in \Sec{sec:crmag}) and here we model this interaction using test-particle (\Sec{sec:crpart}) and MHD (\Sec{sec:crfluid}) simulations. 
The aim is to obtain the spatial distributions of cosmic rays and magnetic fields and then correlate them to test the energy equipartition argument.  

\subsection{Cosmic ray-magnetic field interaction}
\label{sec:crmag}

The Larmor radius $\Lar$ of a cosmic ray particle with a non-dimensional charge $Z$ (charge of the particle divided by the proton or electron charge) and energy $E$ propagating in a magnetic field of strength $B$ is given by
\begin{align}
\Lar \simeq 100 \pc \frac{(E/10^{8} \GeV)}{Z \,\, (B/1 \mkG)} \,\, .
\label{larmor}
\end{align}
Low-energy cosmic ray particles, i.e., particles with Larmor radii less than the correlation length of the magnetic field, propagate diffusively
within the galaxy. This is due to scattering by magnetic fluctuations with scales comparable to the cosmic ray Larmor radius \citep{KP1969, Wentzel1974, Cesarsky1980}. 
The correlation length $\lb$ of small-scale magnetic fields in galaxies as suggested by small-scale dynamo simulations (\Sec{sec:crpart}, \cite{BS13}),
$\lb \simeq (1/3) \, l_0 \approx 30 \pc$, is less than the driving scale of the ISM turbulence, $l_0\approx 100\pc$. For an electron in the ISM magnetic field of around $5 \mkG$, using \Eq{larmor},
we find that the particles with $E \le 10^8 \GeV$ have $\Lar < l_b$.  Further, electrons with energy $E$ in a magnetic field of strength $B$ radiate mostly at the frequency $\nu_{\rm max}$, given by \cite{Webber1980}
\begin{align}
\nu_{\rm max} \simeq 16 \MHz \,\, \frac{B}{1 \mkG} \left(\frac{E}{1 \GeV}\right)^2 \,\, .
\label{numax}
\end{align}
So, for radio frequencies in range $150 \MHz \text{--}1.5 \GHz$ (wavelength $\approx 0.5 \text{--} 20 \cm$), with typical galactic magnetic field of around $5 \mkG$, the electron
energy using \Eq{numax} is in the range $1 \text{--}10 \GeV$. Thus, almost all of the synchrotron generating electrons are propagating diffusively in galaxy disks. 
The protons considered in the equipartition argument are those in the same energy range. The Larmor radius of a $5 \GeV$ cosmic ray particle (proton or electron) in a $5 \mkG$ magnetic field is $10^{-6} \pc$, 
which is much smaller than the correlation length of both the small-scale ($\lb < 100 \pc$) and the large-scale (few $\kpc$ in spiral galaxies)
magnetic fields. The Larmor radius of heavier nuclei is even smaller. Thus, the particles are following the field and it is important to consider the small-scale structure of magnetic field for studying cosmic ray diffusion.
This motivates us to consider cosmic rays as test-particles propagating in a random magnetic field generated by a non-linear small-scale dynamo (\Sec{sec:crpart}).

Cosmic rays via their interaction with magnetic fields also exert pressure on the thermal gas. This affects the gas velocity which in turn affects the magnetic fields. The magnetic field further controls
the cosmic ray propagation and thus it is important to consider the effect of the cosmic ray pressure. The energy density of cosmic rays in the Solar neighbourhood  $E_{\rm cr}$
estimated from the {\it Voyager} and {\it Pioneer} spacecraft data is approximately equal to $1.8 \eV \cm^{-3}$ \citep{Webber1998}. Then the cosmic ray fluid pressure is given by
\begin{align} 
P_{\rm cr} = (\gamma_{\rm cr} - 1)E_{\rm cr},
\label{pcr}
\end{align}
where $\gamma_{\rm cr}$ is the adiabatic index, which is equal to $4/3$ for cosmic rays since it is a relativistic fluid. For $E_{\rm cr}=1.8 \eV \cm^{-3}$, using \Eq{pcr}, we obtain $P_{\rm cr}\simeq4\times10^{-12} \dyn \cm^{-2}$.
For a ISM magnetic field of strength $5 \mkG$, the magnetic field pressure is $P_{\rm B} = B^2/8\pi \simeq 10^{-12} \dyn \cm^{-2}$. Thus, both the pressures are comparable and the cosmic ray pressure significantly affects 
the ISM dynamics. We thus consider cosmic rays as a diffusive fluid in \Sec{sec:crfluid} to account for its pressure contribution.
However, the pressure equality advocated here is with average values and does not necessarily imply that cosmic rays and magnetic fields have the same pressure locally.

\subsection{Cosmic rays as test-particles}
\label{sec:crpart}
\begin{figure*}  \centering
	\includegraphics[width=6.5cm,height=6.5cm]{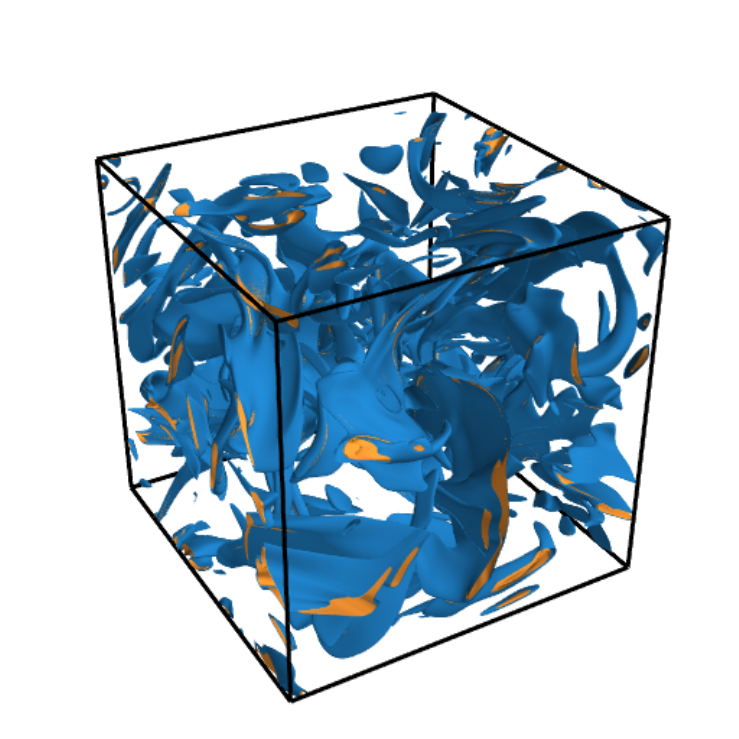}
	\includegraphics[width=7.5cm,height=6cm]{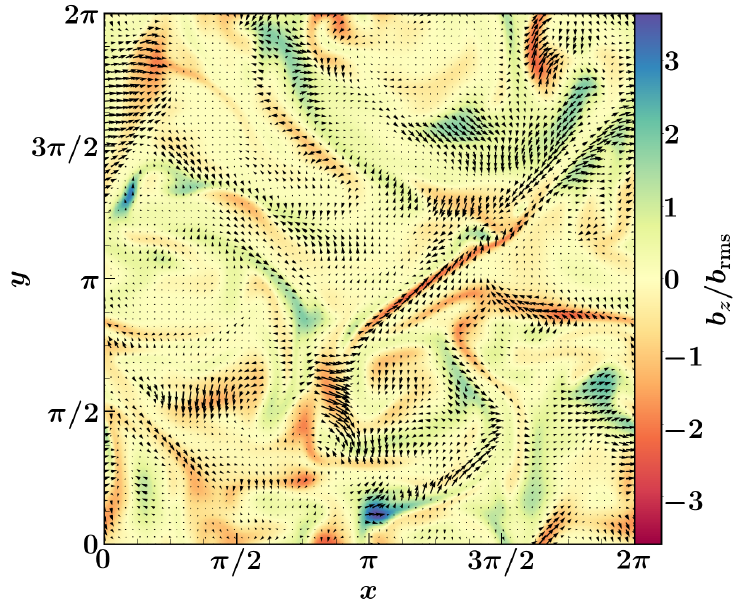}
	\caption{Left panel: Isosurfaces of $b^2/\brms^2 = 3 \, \text{(blue) and} \, 4 \, \text{(yellow)}$ of the magnetic field obtained from a small-scale dynamo simulation. It is intermittent with field
    concentrated in filaments and sheets.
     Right panel: A 2D cut in the xy-plane through the numerical box with vectors for $(b_x/\brms, b_y/\brms)$ and color showing $b_z/\brms$ for the field. The figure shows strong fields in elongated structures.}
	\label{mag}
\end{figure*}

Cosmic rays can be treated as test-particles to study their diffusion in random magnetic fields, especially to understand how the cosmic ray diffusivity depends on the energy of the particle
and properties of the magnetic fields \citep{GJ1999,Casse_et_al2002,DZ2014,Snodin_et_al2016,SSSBW17}. In this work, to obtain random magnetic fields, we numerically solve the equations (\Eq{fdce} - \Eq{fdns})
for the non-linear small-scale dynamo.  For an isothermal gas with equation of state $p_{\rm g} =c_s^2 \rho$, where $p_{\rm g}$ is the gas pressure, $c_s$ is the constant sound speed and $\rho$ is the gas density,
we solve the equations for mass conservation (\Eq{fdce}), magnetic induction (\Eq{fdie}), and the Navier--Stokes equation (\Eq{fdns}) in a periodic box of dimensionless size $(2\pi)^3$ with $(512)^3$ points using the Pencil code 
\footnote{https://github.com/pencil-code}. 
The governing equations are 
\begin{gather}
\frac{\partial \rho}{\partial t} + \nabla \cdot (\rho \vec{u}) = 0 \,\, , \label{fdce} \\
\frac{\partial \vec{b}}{\partial t} = \nabla \times (\vec{u} \times \vec{b}) + \eta \nabla^2 \vec{b} \,\, ,  \label{fdie} \\ 
\frac{\partial \vec{u}}{\partial t} + \vec{u} \cdot \nabla \vec{u} = \frac{-\nabla p_{\rm g}}{\rho} + \frac{ \vec{j} \times \vec{b}}{c\rho} 
+ \nu_{\rm kin} \left(\nabla^2 \vec{u} + \frac{1}{3} \nabla \nabla \cdot \vec{u} + 2 \vec{S} \cdot \nabla \ln \rho \right) + \vec{F} \,\, , \label{fdns}
\end{gather}
where $\vec{u}$ is the velocity, $\vec{b}$ is the magnetic field, $\eta$ is the magnetic diffusivity, $p_{\rm g}$ is the gas pressure, $\vec{j}$ is the current density, $\nu_{\rm kin}$ is the kinematic viscosity, $\vec{S}$ is
the rate of strain tensor, and $\vec{F}$ is the forcing function. The flow is driven by a forcing that is mirror-symmetric, nearly incompressible, and $\delta$-correlated in time, equally at wavenumbers $k = 2\pi/L$ and $k = 2 (2\pi/L)$, 
where $L$ (here $L=2\pi$) is the size of the domain, 
in spectral space \citep{Haugen2004}. The average wavenumber, $\kf$, 
at which the flow is driven is approximately equal to $1.5 (2\pi/L)$ and thus the numerical 
driving scale of turbulence $l_0$ is $2 \pi/\kf$ (physically equal to $100 \pc$). The amplitude of forcing is chosen to ensure that the flow remains subsonic.
For simplicity, we choose $\nu_{\rm kin}  = \eta$, which is chosen to be equal to $10^{22} \cm^2 \s^{-1}$ 
(numerical simulations at the given resolution of $L/512$ resolves the viscous and resistive scales of that order). 
We initialize the box with zero velocity, uniform gas density, and very weak seed Gaussian random magnetic field with mean equal to zero. 
The magnetic field first grows exponentially but then saturates due to the back-reaction by the Lorentz force. 
The correlation length of the saturated small-scale magnetic field $\lb$, calculated using the magnetic power spectrum, is approximately equal to $(1/3) \, l_0$ 
($\lb/l_0$ for the saturated magnetic field does not vary much when the parameters $\nu_{\rm kin}$ and $\eta$ of the simulation are changed within a range \citep{BS13}). 
Here, we present all quantities from the simulations in non-dimensional units.
The magnetic field generated by the small-scale dynamo is spatially intermittent (random field with rare high peaks). \Fig{mag} shows the three-dimensional structure (left-hand panel) and a two-dimensional cut (right-hand panel) of the saturated magnetic field. The magnetic field is concentrated in magnetic structures of various shapes and sizes, mostly sheets and filaments \citep{Wilkin2007}. We expect that a cosmic ray particle
would propagate differently within a magnetic structure and between two such structures. 

\begin{figure*}  \centering
	\includegraphics[width=7.5cm,height=5.5cm]{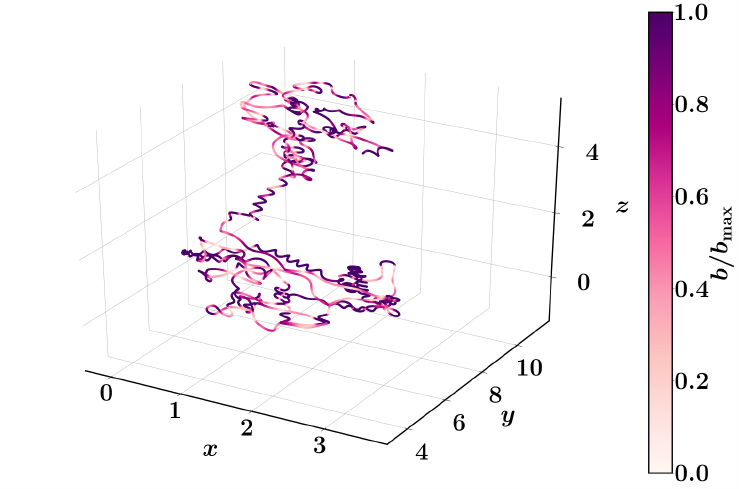}
	\includegraphics[width=7.5cm,height=6.5cm]{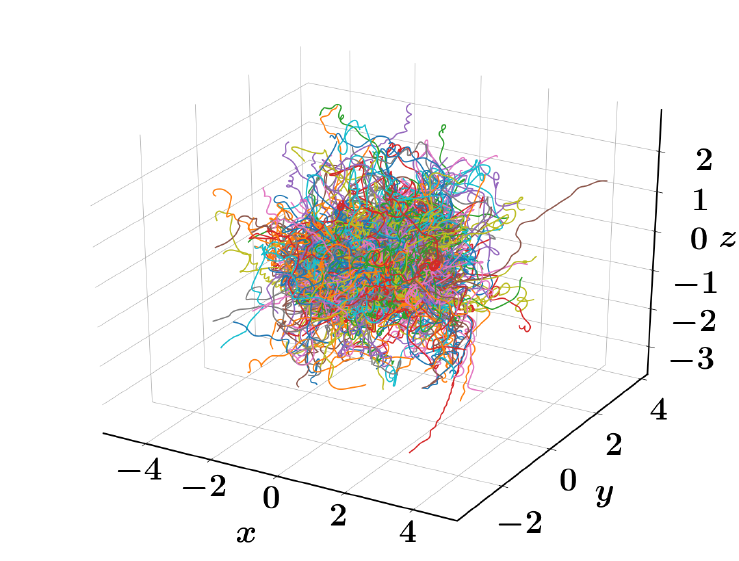}
	\caption{Left panel: Trajectory of a single particle with $\Lar/l_0=0.02$ propagating in the field shown in \Fig{mag}, the color shows the magnetic field strength normalized to its maximum value along the trajectory.
    The particle performs gyrations in strong field regions (following magnetic field lines) but is also scattered in relatively weak field regions.
    Right panel: Trajectories for an ensemble of particles 
     with $\Lar/l_0=0.02$ and same initial spatial location within the numerical domain but random velocity directions (different colored lines is for particles with different initial velocity directions). 
     The particle distribution over large scales in time and length becomes isotropic due to numerous scattering events, which leads to cosmic ray diffusion.}
	\label{traj}
\end{figure*}

The trajectory of a cosmic ray particle is governed by the Lorentz force equation, given by
\begin{equation} \label{lf} 
\oderiv{^2\vec{r}}{t^2} = \frac{v}{\Lar}\,\oderiv{\vec{r}}{t} \times \frac{\vec{b}}{\brms} \,\, , 
\end{equation}
where $\vec{r} (t)$ is the particle position at time $t$, $v$ is the particle speed, $\vec{b}$ is the magnetic field, and $\Lar$ is the Larmor radius of the particle defined with respect
to the root mean square (rms) value of the magnetic field $\brms$ in the domain. $\Lar$ is a measure of the particle energy; in this work we use the non-dimensional parameter $\Lar/l_0$.
Since we are primarily interested in cosmic ray diffusion, we neglect any kind of acceleration of particles. We thus consider
only a time-steady magnetic field (a single snap shot, which is shown in \Fig{mag}). 
Also, we do not consider particle energy losses and thus we model only the proton component of cosmic rays for which the energy loss within the  
confinement time of cosmic rays in galaxies ($\approx 10^{7} \yr$) is negligible. The left-hand panel of \Fig{traj} shows the trajectory of a single particle for $\Lar/l_0 = 0.02$. The particle
gyrates around the magnetic field in structures where the field is strong and is scattered in the region where the field is weak. Over large time- and length- scales for an ensemble of particles (right-hand panel of \Fig{traj}),
this gives rise to diffusion. We solve \Eq{lf} for an ensemble of particles which are placed randomly within the domain with same speed but random velocity directions. We confirm that the particles diffuse by
calculating the time-steady diffusion coefficient \citep{SSSBW17,SSWBS18}. 
Once the diffusion sets in, we calculate the coordinates of each particle modulo the length of the box ($2 \pi$), divide the entire domain into $(512)^3$ smaller cubes and 
count the number of particles in each cube to obtain the particle number density as a function of position and time. Then we average this distribution over a long time to obtain the 
time independent cosmic ray distribution $\ncr$. 

\begin{figure*}
       \centering
        \includegraphics[width=7.5cm,height=7cm]{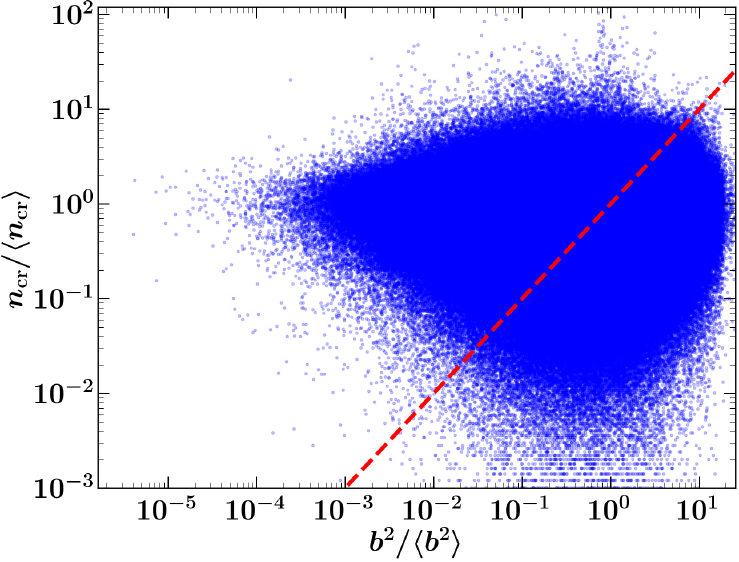} 
         \includegraphics[width=7cm,height=6.5cm]{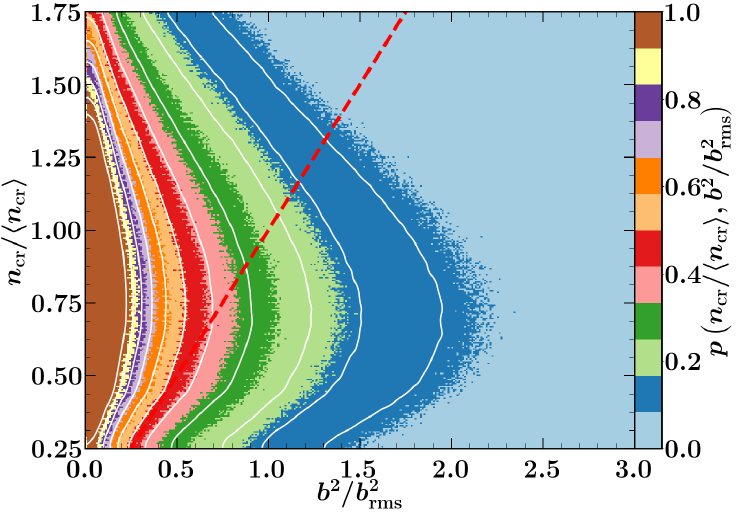} 
	\caption{Left panel: Scatter plot for the normalized number density of cosmic rays $\ncr/\langle \ncr \rangle$ and magnetic fields energy density $b^2/\langle b^2 \rangle$. 
	Right panel: Joint PDF (probability distribution function) of $\ncr/\langle \ncr \rangle$ and $b^2/\langle b^2 \rangle$ with contours showing probability levels $[0.1,0.15,0.2,0.3,0.4,0.5,0.6,0.7,0.8,0.9,0.99]$
	 for a range in cosmic ray number density $(0.25 \le \ncr/\langle \ncr \rangle \le 1.75)$ and 
	 magnetic field energy density $(0.0 \le  b^2/\langle b^2 \rangle \le 3.0)$ where most points ($\approx 95\%$) of the left-hand panel lie.
	 The red dashed line shows a one-to-one correlation between the two quantities in both panels. 
	The cloud of points in both panels confirms that the cosmic ray number density and magnetic field energy density are uncorrelated in test-particle simulations.}
        \label{tpcorr}
\end{figure*}

The cross-correlation coefficient between cosmic rays and magnetic fields is calculated using
\begin{equation}\label{crosscorr}
	  C(\ncr,b^2) = 
	  \frac{\langle \ncr b^2 \rangle 
	  	- \langle \ncr \rangle \langle b^2 \rangle}{\sigma_{\ncr} \sigma_{b^2}} \,\, ,
\end{equation}
where $\langle \cdots \rangle$ and $\sigma$ implies average and standard deviation of quantities over the entire domain. The cross-correlation coefficient $C$ is approximately
equal to zero for various particle energies ($\Lar/l_0$). This confirms that the two quantities are uncorrelated at scales less than the driving scale of the turbulence $l_0$ (here approximately the size of the box). 
The correlation remains close to zero even when a regular large-scale field and pitch
angle scattering due to small-scale magnetic fluctuations (which are not resolved in the nonlinear small-scale dynamo simulation \citep{SSWBS18}) are included.
The left-hand panel of \Fig{tpcorr} shows a scatter plot of the cosmic ray number density and magnetic field energy density, which further confirms the lack of correlation between two quantities. 
The right-hand panel of \Fig{tpcorr} shows the joint probability distribution function (PDF) of the cosmic ray number density and magnetic field energy density where the majority of points lie.
This too demonstrates that the most points lie away from the red dashed line which shows a one-to-one correlation between two quantities.

The left-hand panel of \Fig{tpcorr} shows 
cosmic ray number densities which are significantly higher and lower than the mean value. This suggests an inhomogeneous cosmic ray distribution. Due to a finite number of particles, the initial distribution
in such simulations is not completely isotropic and homogeneous, this makes the final distribution inhomogeneous \footnote{In an ideal situation, a perfectly isotropic and homogenous 
distribution would always remain isotropic and homogenous in a static magnetic field due to the Liouville's theorem.}. These localized areas of high cosmic ray density are the regions
with magnetic bottle traps and this is confirmed in numerical simulations by looking at particle trajectories in the vicinity of a trap \citep{SSWBS18}. The presence of such traps enhances cosmic
ray density at certain locations (or reduces at others) but those are not necessarily the strong magnetic field (or the weak magnetic field) regions. 
Furthermore, trapping might change the proton to electron ratio $K$. Both protons and electrons
would spend more time in magnetic traps (as compared to non-trapping regions), but electrons, since they lose energy, get slower and are trapped for a longer time as compared to protons. 
This would change the factor $K$ (also see the point (2) in \Sec{sec:res}). Moreover, in such a situation $K$ would not be a fixed constant and will depend on the location and energy of the particle. 
Having said that, such traps are only of relevance at small scales (less than the correlation length of the small-scale magnetic field, $l_b \simeq (1/3) \, l_0 \approx 30 \pc$) 
and may not affect the conclusions at larger $\kpc$ scales.

\subsection{Cosmic rays as a diffusive fluid in MHD turbulence}

\label{sec:crfluid}
To consider the effect of the cosmic ray pressure, we solve the MHD equations
using a two-fluid model: gas with adiabatic index $\gamma_{\rm g} = 5/3$ (a non-relativistic fluid)
and cosmic rays with adiabatic index $\gamma_{\rm cr} = 4/3$ (a relativistic fluid).
For an isothermal gas with equation of state $p_{\rm g} =c_s^2 \rho$, where $p_{\rm g}$ is the gas pressure, $c_s$ is the constant sound speed and $\rho$ is the gas density,
we solve equations for mass conservation (\Eq{fdce}), magnetic induction (\Eq{fdie}), Navier-Stokes equation (\Eq{fdns}) but now also including the cosmic ray pressure (i.e. the first term in the right-hand side is 
modified to $-\nabla (p_{\rm g} + p_{\rm cr})/\rho$), and cosmic ray advection-diffusion (\Eq{cradvectdiff1} and \Eq{cradvectdiff2})
in a box of dimensionless size $(2\pi)^3$ (periodic boundary condition for $y$ and $z$, 
stress-free normal field boundary condition with energy density of cosmic rays $e_{\rm cr} = 0$ at boundaries $x=0,2\pi$) with $256^3$ points. The forcing function and its parameters for a turbulent driving
in the Navier-Stokes equation is same as described in \Sec{sec:crpart}. The cosmic ray advection-diffusion equation is 
\begin{align}  
\frac{\partial e_{\rm cr}}{\partial t} + \nabla \cdot (e_{\rm cr} \vec{u}) + p_{\rm cr} \nabla \cdot u = - \nabla \cdot \vec{F_{\rm cr}} + Q_{\rm cr} \,\, ,
\label{cradvectdiff1}
\end{align}
where $e_{\rm cr}$ is the cosmic ray energy density, $p_{\rm cr} = (\gamma_{\rm cr} - 1)e_{\rm cr}$ is the cosmic ray pressure, 
$Q_{\rm cr}$ is the cosmic ray energy source by which cosmic rays are injected uniformly throughout the volume at a constant rate (note that the cosmic rays are lost via the boundaries in $x$ direction). 
The cosmic ray flux $\vec{F_{\rm cr}}$ is defined via the equation
\begin{align}
 \tau \frac{\partial F_{{\rm cr}i}}{\partial t} = -\kappa_{ij} \nabla_j e_{\rm cr} - F_{{\rm cr}i} \,\, ,
\label{cradvectdiff2}
\end{align}
where $\tau$ is the cosmic ray
flux correlation time and $\kappa_{ij}$ is the diffusion coefficient. $\kappa_{ij} = \kappa_{\perp} \delta_{ij} + (\kappa_{\parallel} - \kappa_{\perp}) \, \uvec{b}_i \uvec{b}_j$ is 
written in terms of the parallel $\kappa_{\parallel}$ and perpendicular $\kappa_{\perp}$ diffusion coefficients  ($\uvec{b}_i$ is the unit vector along $i$-axis).
Here, we solve the telegraph equation (\Eq{cradvectdiff2}) to study cosmic ray diffusion instead of the usual diffusion equation \citep{SBMS06,RSSS18}.

\begin{table}
	\centering
	\caption{Non-dimensional parameters used to solve the MHD and cosmic ray fluid equations and their corresponding ISM values.}
	\label{tablecr}
	\begin{tabular}{ccc} 
		\hline
		Parameter  & Numerical value & ISM value \\
		\hline
                  $\gamma_{\rm g}$ & $5/3$ &  $5/3$\\
                  $\gamma_{\rm cr}$ & $4/3$ & $4/3$\\
                  $\kf$ & $1$ and $2$ & $100 \pc$ and $50 \pc$\\
                  $\nu_{\rm kin}$ & $2\times10^{-3}$ & $ 2 \times 10^{23} \cm^2 \s^{-1}  $  \\
                  $\eta$ & $1\times10^{-3}$ & $1 \times 10^{23} \cm^2 \s^{-1} $ \\
                  $\kappa_{\parallel}$ & $3\times10^{-1}$ & $ 3 \times 10^{25} \cm^2 \s^{-1}$  \\
                  $\kappa_{\perp}$ & $0$ & $0$ \\ 
                  $\tau$ & $3\times10^{-1}$ & $ 0.3 \Myr$\\
                  $Q_{\rm cr}$ & $0.001, 0.005, 0.01$ & $ (0.001, 0.005, 0.01) \times 10^{-26} \erg \cm^{-3} \s^{-1} $\\
		\hline
	\end{tabular}
\end{table}

\begin{figure*} \centering
	\includegraphics[width=10cm]{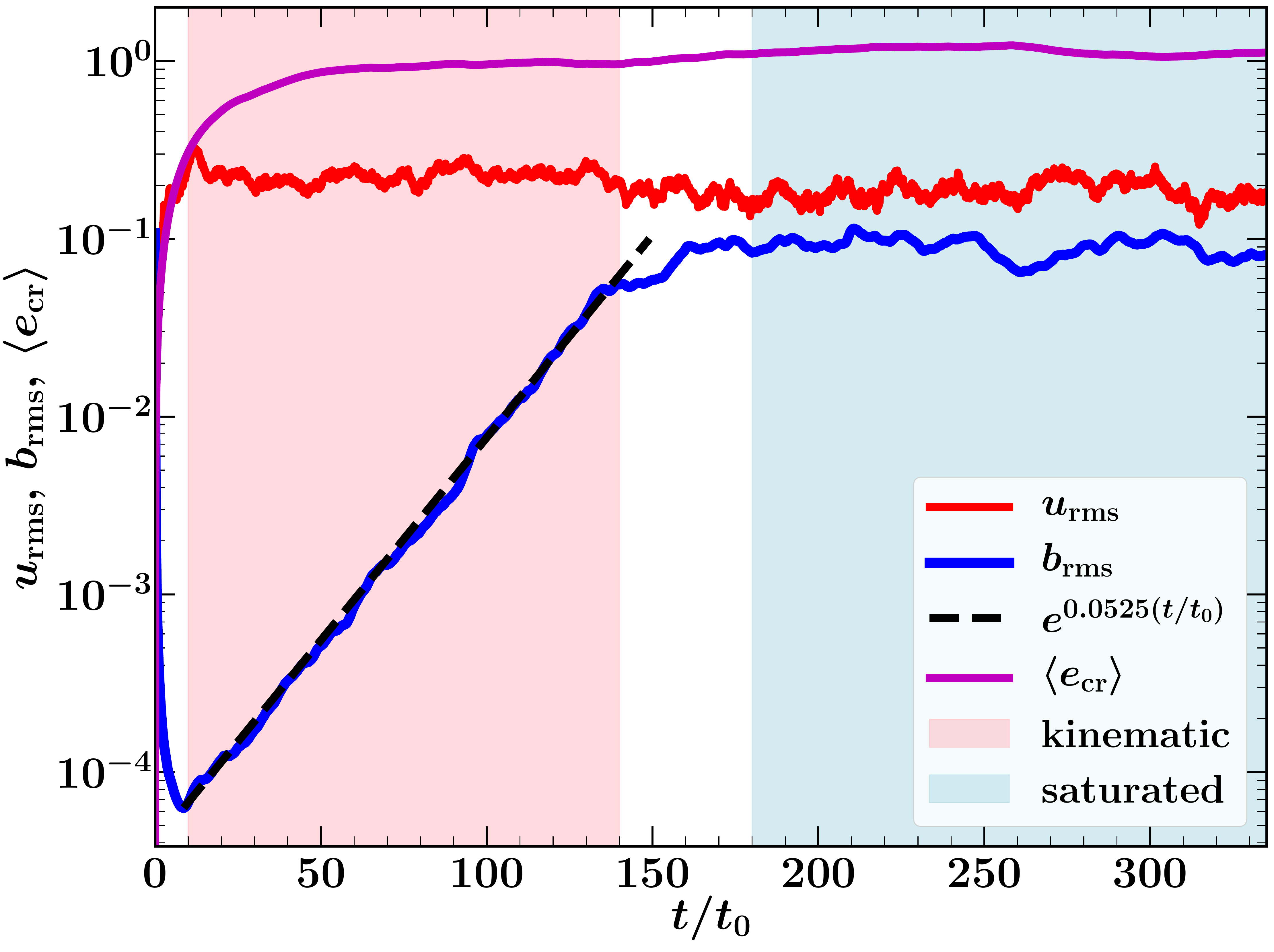}
	\caption{Time evolution of the rms velocity field $u_{\rm rms}$ (red) , rms magnetic field $b_{\rm rms}$ (blue), and mean cosmic ray energy density $\langle e_{\rm cr} \rangle$ 
	(magenta) as a function of normalized time $t/t_0$, where $t_0$ is
	the eddy turnover time, for the case where $\langle e_{\rm cr} \rangle > b^2_{\rm rms}$. 
	The magnetic field decreases until it transforms into an eigenfunction of the induction equation and then it grows exponentially (kinematic stage, light pink).
	 The magnetic field finally saturates (saturated stage, light blue) due to the back reaction on the velocity flow by Lorentz forces. The cosmic ray energy density saturates faster than 
	 the magnetic field. }
	\label{tscr}
\end{figure*}

We solve these equations using the Pencil code \citep{SBMS06} for parameters given in Table~\ref{tablecr}
\footnote{For a $\GeV$ cosmic ray particle in a $\mkG$ magnetic field, the parallel cosmic ray diffusivity $\kappa_{\parallel}$ in the ISM is approximately $10^{28} \cm^{2} \s^{-1}$ \citep{BBDP90}. 
This number is not yet accessible in numerical simulations where the turbulence is driven at the box scale of a physical size $100 \pc$. So, we also decrease
the magnetic field diffusivity $\eta$ in our simulations. The magnetic field in the ISM mostly diffuses via turbulent diffusion with the diffusivity of the order of $10^{26} \cm^2 \s^{-1}$ \citep{Shukurov2004}.
Thus, in our numerical simulations, we chose $\kappa_{\parallel}$ and $\eta$ such that the ratio of these two terms is $\kappa_{\parallel}/\eta \approx 10^{-2}$. \label{f1}}. 
All velocities are in units of the sound speed $c_s$, densities are in units of the initial gas density $\rho_0$,
lengths are in units of the box size $L=2\pi$ (so that the smallest wavenumber is $k_{1}= 2\pi/L$), time is in units of the eddy turn over time $t_0 = 1/u_{\rm rms} \kf$,
the magnetic field in units of $\left(4 \pi \rho_0 c_{s}^2\right)^{1/2}$ and all the
diffusivities are in units of $c_s/k_{1}$. All other units can be derived from these basic units. The unit of the cosmic ray source term is $Q_{\rm cr}$ is $\rho_0 c_s^3 k_1$.
We select $\kappa_{\perp} = 0$ for simplicity.  
The value of $\tau$ is chosen such that the maximum speed for signal propagation is of the order of $(\kappa_{\parallel}/\tau)^{1/2}$. This is done
to capture the initial non-diffusive or ballistic phase of cosmic rays in random magnetic fields \citep{RSSS18}.

There are a few differences between the ISM parameters in our simulations (Table~\ref{tablecr}) and those estimated for the ISM from observations or existing models.
First, it is inferred from the observed radio polarization gradients that the ISM turbulence is quite compressive with a Mach number around 1
\citep{Gaensler2011}, while our simulations are at a very low Mach number ($\sim 0.1$). Second, using kinetic theory
arguments, the ratio of resistivity to viscosity in the ISM can be estimated to be of the order of $10^{11}$ \citep{BS2005},
however, for our simulations, it is of the order of one. Furthermore, the cosmic ray diffusivity is also chosen to be smaller than the reported
value of $10^{28} \cm^{2} \s^{-1}$ \citep{BBDP90}, a number obtained from the abundance ratio of the radioactive and stable isotopes of Beryllium in cosmic rays. All
diffusive effects (viscosity, resistivity and cosmic ray diffusivity) are much lower than those estimated in the ISM due to the limited numerical resolution.
These estimated parameters might be important for comparing simulations with the observations but do not affect the physics of cosmic ray -- magnetic field interaction 
(additional pressure contribution due to cosmic rays) which we aim to capture in these numerical experiments.

\begin{figure*}
       \centering
         \includegraphics[width=7.5cm,height=7cm]{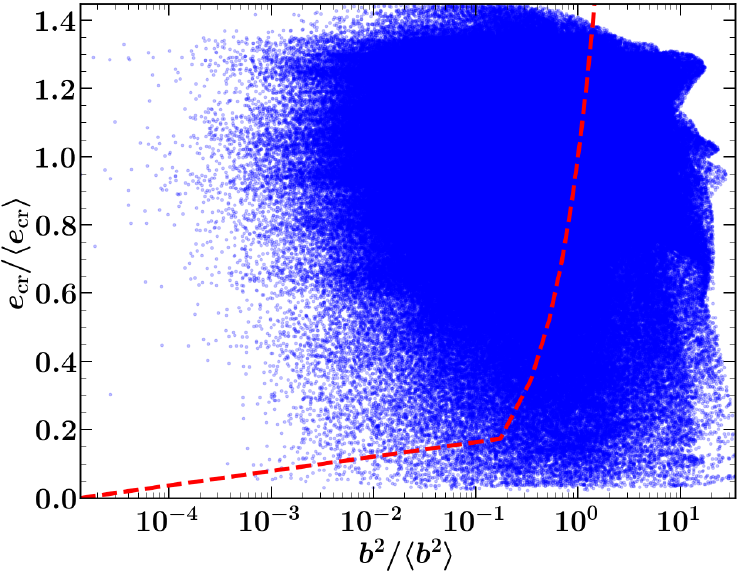} 
          \includegraphics[width=7cm,height=6.25cm]{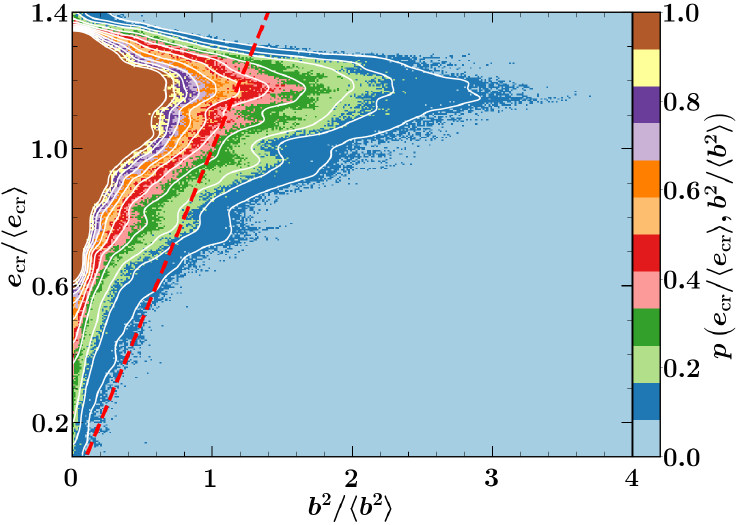} 
	\caption{Left panel: Scatter plot for the normalized energy densities of cosmic rays $e_{\rm cr} /\langle e_{\rm cr}  \rangle$ and magnetic fields  $b^2/\langle b^2 \rangle$ for the case
	with $\langle e_{\rm cr}  \rangle \approx \langle b^2 \rangle$ at the box scale of a physical size $100 \pc$ (parameters chosen to ensure the same). 
	Right panel: Joint PDF of $e_{\rm cr} /\langle e_{\rm cr}  \rangle$ and $b^2/\langle b^2 \rangle$ with contours showing probability levels $[0.1,0.15,0.2,0.3,0.4,0.5,0.6,0.7,0.8,0.9,0.99]$
	 for a range in cosmic ray number density $(0.1 \le e_{\rm cr} /\langle e_{\rm cr}  \rangle \le 1.4)$ and 
	 magnetic field energy density $(0.0 \le  b^2/\langle b^2 \rangle \le 4.0)$ where most points ($\approx 95\%$) of the left-hand panel lie.
	 The red dashed line shows the perfect correlation between two quantities in both panels (note the semi-log scale in the left-hand panel). 
	 Even though both energy densities are equal when averaged over the size of the domain, locally they are uncorrelated.}
        \label{fluidcorr}
\end{figure*}

We choose three different values for the cosmic ray source term $Q_{\rm cr}$ to have following three different cases in the saturated stage:
$\langle e_{\rm cr} \rangle  < b^2_{\rm rms}, \, \langle e_{\rm cr} \rangle  \approx  b^2_{\rm rms}, \, \langle e_{\rm cr} \rangle > b^2_{\rm rms}$.
A uniform gas density and a weak (in rms sense) Gaussian random magnetic field with zero mean is initialized within the numerical domain.
The velocity and cosmic ray energy density are both initialized to zero and with time they both become non-zero due to continuous forcing $\vec{F}$ 
in the Navier--Stokes equation (\Eq{fdns}) and the cosmic ray source term $Q_{\rm cr}$ in the advection--diffusion equation (\Eq{cradvectdiff1}), respectively.
\Fig{tscr} shows the evolution of the root mean square (rms) velocity field, 
rms magnetic field, and mean cosmic ray energy density. The magnetic field first decays until it transforms into a growing eigenfunction of the induction equation
\footnote{The seed magnetic field is initialized to be a Gaussian random magnetic field, which is not a solution of the induction equation. Thus, the field decays initially as shown in \Fig{tscr}.}
and then it increases exponentially (referred to as kinematic stage). 
Finally, the magnetic field becomes strong enough and the Lorentz force reacts back on the velocity flow saturating 
the magnetic field (referred to as saturated stage). For all three cases, we find that the cross correlation between cosmic rays and magnetic fields 
(using \Eq{crosscorr}) is very close to zero. Thus, even after including the effect of cosmic ray pressure on the thermal gas, the cosmic rays and magnetic fields
are not correlated at the scales less than the driving scale of the turbulence ($l_0 \simeq 100\pc$ or equivalently the numerical domain size in simulations). The left-hand panel of \Fig{fluidcorr} shows the scatter plot of the
cosmic ray and magnetic field energy densities over the entire domain for the case where $\langle e_{\rm cr}  \rangle \approx \langle b^2 \rangle$ in the saturated state. The form
of the plot confirms that the two quantities are not correlated. 
Furthermore, the right-hand panel of \Fig{fluidcorr} shows the joint PDF between the two quantities for most points in the domain, which too shows lack of correlation.
Thus, even when both energies are equal over the scale of the domain (by construction here),
they are not correlated locally, so that equipartition is not valid.

\section{Conclusions, discussion and future directions of research}
\label{sec:con}

Synchrotron emission depends on two quantities: cosmic ray electron number density and magnetic fields, so that assuming a relation between two quantities provides information about magnetic fields from synchrotron observations in star-forming galaxies. Thus, the energy equipartition argument is a convenient assumption.
However, it is not a physical law and thus it is important to test its validity. After describing the method and theoretical arguments to expect the energy equipartition in star-forming galaxies, this paper discusses its validity using observational results and numerical simulations.

There is convincing observational evidence that equipartition between the energy densities of total magnetic fields and total cosmic rays is valid in normal star-forming galaxies, globally as well as locally at spatial scales larger than a certain threshold, possibly related to the mean propagation length of the cosmic rays of $\approx 1 \kpc$, and can be safely used to estimate the total magnetic fields strength from the total synchrotron intensity. If the synchrotron-radiating CREs have lost a significant fraction of their energy, e.g. in dense regions of the interstellar medium or in galaxy halos, a corrected proton/electron ratio has to be used in the formula. Equipartition is not valid in starburst regions or in ultraluminous infrared galaxies. The equipartition condition may also fail in dwarf galaxies \citep{Filho2019}.

Future observations should investigate the threshold scale between the regime of equipartition and non-equipartition and its dependence on galaxy properties. The exponent of the radio--infrared correlation within galaxies should be measured at different scales, to fix the transition from super-linear to sub-linear. As the threshold scale is possibly related to the average propagation length of CR(E)s, a spatial resolution down to about 100\,pc for a sample of galaxies is needed, which calls for high angular resolution (1--2\,arcsec) in galaxies at 10--20\,Mpc distances. Present-day radio telescopes allow us to observe only the nearest and brightest galaxies with such a resolution. Increasing the galaxy sample will have to wait for the Square Kilometre Array (SKA) \cite{Beck2015b}.

More and better tests of equipartition with help of $\gamma$-rays require sensitive observations of many nearby galaxies with different star-formation rates. ESA's e-ASTROGAM satellite project plan is promising \cite{Angelis2018}.

Using both, the test-particle simulations in realistic (small-scale dynamo generated) random magnetic fields and the MHD simulations, we show that the cosmic ray energy density (or equivalently number density) 
is not correlated to the magnetic field energy density at scales smaller than the driving scale of the turbulence ($\approx 100 \pc$ in spiral galaxies).
This scale can be regarded as the lower limit for the scale beyond which equipartition is valid.
In test-particle simulations, the particles sample the small-scale structure of random magnetic fields which is an important ingredient (magnetic power at smaller scale) 
to justify energy equipartition assumption via the confinement argument
(second argument in \Sec{sec:why}). In MHD simulations, the effect of cosmic ray pressure is also included. This is considered to test the local pressure equality between cosmic rays and magnetic fields. 
The results from both kinds of simulations (\Fig{tpcorr} and \Fig{fluidcorr}) suggest a lack of energy equipartition at smaller scales. 
This is in agreement with the observational results (\Sec{sec:equigal} and \Sec{sec:dev}). However, our simulations do not exclude that energy equipartition is valid at larger scales.

Cosmic rays in our MHD simulations only exert pressure on the thermal gas but it can also heat up the medium. 
Cosmic ray scatter off waves excited by the streaming instability \citep{KP1969,Wentzel1974,Skilling71}
and then stream at the Alfv\'en speed down their pressure gradient. The excited waves are damped and their energy is deposited in the ISM, which heats it up. 
This could be modelled in our numerical simulations (a stable numerical scheme for the cosmic ray streaming is discussed in \cite{Sharma2010,TP19}). 
The cosmic ray streaming is particularly important for the launching of galactic winds \citep{RYZ17,Zweibel17} and the Parker instability \citep{HZ18}.
Another extension of the numerical work would be to include cosmic rays in comparatively larger scale 
multiphase ISM simulations (domain size of a few $\kpc$)  where the turbulence is driven by supernova explosions \citep{Gent_et_al12013,Li15,Kim2017}.
The cosmic rays would then accelerate in supernova shocks and diffuse away from their sources. One might recover equipartition at large scales in such studies.

\authorcontributions{Conceptualization, A.S. and R.B.; methodology,  A.S.; software, A.S.; validation, A.S. and R.B.; formal analysis,  A.S. and R.B.; investigation, 
 A.S. and R.B.; resources,  A.S. and R.B.; data curation,  A.S. and R.B.; writing--original draft preparation,  A.S. and R.B.; writing--review and editing,  A.S. and R.B.; visualization,  A.S. and R.B.}

\funding{This research received no external funding.}

\acknowledgments{A.S. thanks Anvar Shukurov, Paul  Bushby, Toby Wood, Andrew Fletcher, and Torsten En{\ss}lin for several useful discussions. 
We thank Ellen Zweibel, Luke Chamandy, Elly M. Berkhuijsen and Marita Krause for carefully reading the manuscript. We thank Aritra Basu and Elly M. Berkhuijsen 
for providing \Fig{fig:RIC} and \Fig{fig:MW} respectively.}

\conflictsofinterest{The authors declare no conflict of interest.} 




\externalbibliography{yes}
\bibliography{equ}



\end{document}